\documentclass[a4paper] {article}
\usepackage{graphicx}

\begin{document}
\parindent=0pt
\parskip=0.3cm

\def \e{\hbox{e}}
\def \half{\textstyle {1 \over 2}}
\def \exp{\hbox{exp}}


\null
\vskip 1.5cm
\centerline{FINGERS OF GOD}
\vskip 0.6cm
\centerline{A CRITIQUE OF REES'S THEORY OF PRIMORDIAL GRAVITATIONAL RADIATION}
\vskip 1.5cm

\centerline {J.C. Jackson\footnote{Present address: Division of Mathematics, School of Computing, Engineering and Information Sciences, Northumbria University, Newcastle NE1 8ST, UK}}
\centerline {Royal Greenwich Observatory}
\centerline {Herstmonceux Castle}
\centerline {Sussex}
\centerline {UK}
\vskip 1.5cm

\centerline {(Received 1971 November 12)}
\vskip 1.5cm

\centerline{\bf SUMMARY}
\vskip 0.3cm

Very long wavelength universal gravitational waves cannot now produce in
clusters of galaxies velocity dispersions greater than that which these
systems would possess if they were expanding with the Universe, if the
Universe is not younger than $10^{10}$ yr and Hubble's constant is not less
than 50 km/sec/ Mpc. A diagram shows that actual velocity dispersions
are significantly greater than this limit.
\newpage

\centerline{INTRODUCTION}

The persistently negative results of searches for enough ordinary matter
in clusters of galaxies to account for their large velocity dispersions will
probably stimulate interest in `radical' solutions of this problem. I shall
discuss here one such proposal, namely Rees's hypothesis (1971) that these
systems are interacting with very long wavelength gravitational waves. Before
doing so, however, I shall present a diagram which shows very clearly what
I consider to be the essential feature of this phenomenon; it is this feature
which cannot be explained by Rees's theory in its present form.

Fig. 1 shows the spatial distribution of all galaxies with right ascensions
in the range 10-11 hr, and known positive radial velocities (relative to the
local group) less than 2000 km s$^{-1}$. The `distance' $D$ used in this plot
is defined as velocity/100 km/sec/Mpc. The galaxies appear to fall into long
chains or cigar-shaped configurations, all pointing at the Earth. Unless one
is prepared to assign to the Earth a very special place in the Universe, one
must conclude that $D$ is not a good distance indicator, and that in reality the
galaxies exist in roughly spherical configurations whose internal velocity
dispersions are several times that which would be observed if these systems
were expanding with the Universe. (I shall call the latter the internal Hubble
velocity.) It is this aspect of these small groups, rather than their mass
discrepancy, which is odd, as in many cases the virial theorem velocity
corresponding to the visible mass is considerably smaller than the internal
Hubble velocity. I shall show that gravitational waves could at most produce
a velocity dispersion equal to the internal Hubble velocity.

The groups apparent in Fig. 1 coincide with members of de Vaucouleurs' complete
list (1971) of small groups within 17 Mpc of the local group, which was compiled
using distance indicators other than velocity. An analysis of this list gives a 
mean value of the ratio internal velocity dispersion/internal Hubble velocity $\sim$ 5.
Although the precise value of this number is affected by uncertainties in group
membership, diagrams such as Fig. 1 convince one that the groups are real and
that this ratio is significantly greater than unity.

The apparent instability exhibited by de Vaucoulers' groups is not particularly
violent; their crossing times are longer than $10^9$ yr (Rood, Rothman \& Turnrose 1970).
Systems with much shorter crossing times are known (see for example Burbidge \& Sargent
1970); however, following Rees I shall restrict my discussion to groups with long
crossing times.
\vskip 0.6cm

\centerline{GRAVITATIONAL WAVES}

The interaction of a plane gravitational wave with a cluster of galaxies can
be solved exactly in the linearized version of general relativity (for a suitable
theoretical background see Weber 1961). Such a wave, moving along the x-axis with
angular frequency $\omega$, is described by the metric

$$
g_{\mu\nu}=\eta_{\mu\nu}+h_{\mu\nu}
$$

where $\eta_{\mu\nu}$ is the Minkowski metric, and the first order perturbation
$h_{\mu\nu}$ is given by

$$
h_{\mu\nu}=Ae_{\mu\nu}\exp\left[i\omega\left(t-{x \over c}\right)\right]
$$

($\mu,\nu = 0, 1, 2, 3$ indicates $t, x , y, z$ components respectively).

$A$ is the amplitude of the wave, and for a fully polarized one the polarization
tensor $e_{\mu\nu}$ can be

\begin{figure}
\begin{center}
\includegraphics[width=15.5cm] {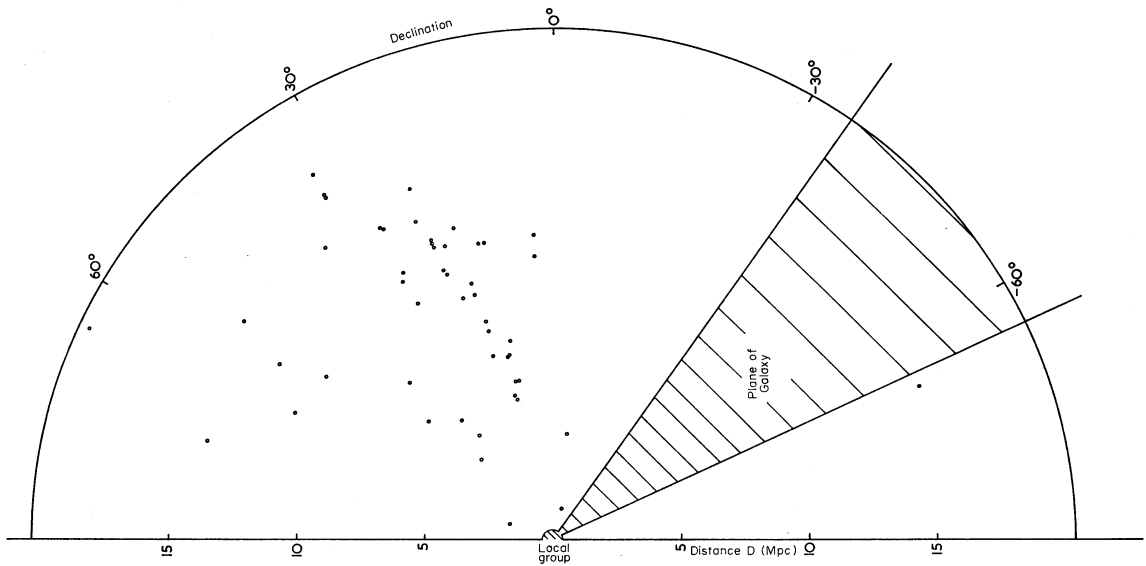}
\caption{The apparent space distribution of nearby galaxies with 10h $\leq RA\le$ 11h,
assuming radial velocity to be a good distance indicator.}
\end{center}
\end{figure}

written as

$$
e_{\mu\nu}={1 \over \surd{2}}(\delta_\mu^{~2}\delta_\nu^{~2}-\delta_\mu^{~3}\delta_\nu^{~3})
$$

where $\delta_\rho^{~\sigma}$ is the Kronecker $\delta$ symbol. Consider now the interaction
of this radiation with a system of $N$ point particles, with masses $m_i$, and position
vectors $(x_i, y_i, z_i)$ and velocities $(u_i, v_i, w_i)$ in centre of mass coordinates;
the gravitational interaction between these particles will be neglected. (The masses $m_i$
could be omitted from this discussion; the advantage of including them is that the resulting
mass weighted quantities are readily available for many systems.) Their accelerations relative
to the centre of mass are related to the Riemann tensor $R_{\mu\nu\rho\sigma}$ via the equation
of geodesic deviation:

$$
\ddot x_i=0
$$
$$
\ddot y_i=R_{0202}y_i=-\half\ddot h_{22}y_i
=\half\omega^2Ae_{22}y_i\exp\left[i\omega\left(t-{x \over c}\right)\right]
$$
$$
\ddot z_i=R_{0303}z_i=-\half\ddot h_{33}z_i
=\half\omega^2Ae_{33}z_i\exp\left[i\omega\left(t-{x \over c}\right)\right]
$$

where a dot indicates a time derivative. If, as in the situation envisaged by Rees,
the particles do not suffer any sensible change of position during the course of an
oscillation, and the only appreciable velocities are those induced by the wave,
these equations integrate to

$$
u_i=0
$$
$$
v_i=\half\omega Ae_{22}y_i\exp\left[i\omega\left(t-{x \over c}\right)-i\pi/2\right]
$$
$$
w_i=\half\omega Ae_{33}z_i\exp\left[i\omega\left(t-{x \over c}\right)-i\pi/2\right]
$$

Let $M$ be the total mass of the system, and define the velocity dispersion V by

$$
MV^2=\hbox{time average of }\left[\sum_i m_i(|u_i|^2+|v_i|^2+|w_i|^2)\right].
$$

Then

$$
MV^2
=\textstyle{1 \over 16}\omega^2A^2\sum_i m_i(y_i^2+z_i^2)
\sim\textstyle{1 \over 16}\omega^2A^2\times \textstyle{2 \over 3}\sum_i m_i(x_i^2+y_i^2+z_i^2)
$$

so that

$$
V^2\sim\textstyle{1 \over 24}\omega^2A^2R_I^2
$$
where $R_I$ is the radius of gyration of the system about its centre of mass:

$$ 
MR_I^2=\sum_i m_i(x_i^2+y_i^2+z_i^2).
$$

We are concerned with wavelengths $\lambda<<$ radius of the Universe,
and the effective mass density of this radiation is (Isaacson 1968)

$$
\rho_G={A^2\omega^2 \over 64\pi G}.
$$

Thus in terms of $\rho_G$, we have

$$
{V \over R_I}=(\textstyle{8 \over 3}\pi G\rho_G)^{1/2}.
\eqno(1)
$$

For a truly plane wave, this expression is valid for all values of $\lambda$;
Rees notes that it would be unreasonable to suppose that the planeness extends
over regions much greater than $\lambda$, so that if $\lambda<R_I$ this expression
should be reduced by a factor $\lambda/R_I$. However, when $\lambda<<R_I$, the
gravitational waves will act as a smoothed out medium of density $\rho_G$ for the cluster
as well as for the Universe, and the velocity dispersion (1) is precisely that 
which the ordinary gravity of such a medium (if static) would produce in the system.
Thus equation (1) is possibly valid for all values of $\lambda$, although the last
point is not clear as the medium would be expanding on a timescale comparable with 
the time taken by members to cross the system. Equation (1) certainly gives an upper
limit for Rees's mechanism whatever the value of $\lambda$. Note that the wave has
no associated gravitational redshift to augment the effect, as $h_{00}=0$.

The results of this calculation agree with Rees' order of magnitude estimates;
however, it is important to establish that $R_I$ is the correct radius to use
in this context, as radii defined in other ways differ considerably in magnitude.
For example, the radius $R_\Omega$, defined so that $-GM^2/R_\Omega$ is the
gravitational potential energy of the system, appears in applications of the
virial theorem to groups (see for example Rood, Rothman \& Turnrose 1970),
and is typically larger than $R_I$ by a factor of 3. The use of $R_\Omega$ in
equation (1) would therefore considerably overestimate the effect. $V$ is
maximized by maximizing $\rho_G$, i.e. by considering a universe in which almost
all the mass-energy is in the form of radiation. The observed deceleration of the
Universe would then be due entirely to this radiation, and the deceleration
parameter $q_0$ would be given by (Sandage 1961)

$$
q_0=\textstyle{8 \over 3}\pi G\rho_G H_0^{-2}
$$

where $H_0$ is Hubble's constant. In this case, equation (1) becomes

$$
{V \over R_I}=q_0^{1/2}H_0.
\eqno(2)
$$

It is easy to show that under these conditions $q_0$ is almost certainly less
than unity. Thus we see that gravitational waves could at most produce a velocity
dispersion in a system equal to that which would be produced by the Hubble
expansion. If the observed values of $V/R_I$ really were of order $H_0$, rather
than $5H_0$, it is doubtful that astronomers would consider that clusters of
galaxies presented any problems. The upper limit on $q_0$ is derived as follows;
let $t_0$ be the present age of the Universe; then the inequalities

$$
t_0\geq 10^{10}\hbox{ yr}~~~~~~~~~~~~
H_0\geq 50\hbox{ km/sec/Mpc}
$$

are almost certainly true. Thus

$$
H_0t_0\geq\half
\eqno(3)
$$

which, for a universe filled with radiation, implies

$$
q_0\leq 1.
$$
(The value $q_0=25$ required to produce the observed values of $V/R_I$ would
need $H_0\leq 17$ km/sec/Mpc if $t_0\geq 10^{10}$ yr.)

I shall conclude by noting that another recently suggested `radical' solution
of this problem (Jackson 1970; see also Forman 1970; Paal 1964), which attributes
the velocity dispersions to a negative cosmological constant, suffers from similar,
but less severe, problems. In fact this theory also gives rise to equation (2);
however, in the appropriate universe, inequality (3) leads to

$$
q_0\leq 5.4
$$
and thus
$$
{V \over R_I}<2.3.
$$
\newpage

\centerline{ACKNOWLEDGMENTS}

It is a pleasure to thank Dr D. Lynden-Bell for several stimulating conversations
on this subject.
\vskip 0.3cm

{\it Royal Greenwich Observatory, Herstmonceux Castle, Sussex}
\vskip 0.9cm

\centerline{REFERENCES}
\vskip 0.3cm

{\obeylines\parskip=0pt
Burbidge, E. M. \& Sargent, W. L. W., 1971. Pont. Acad. Scient. Scripta Varia 35 (Vatican
{\parindent=0.6cm Conference on the Nuclei of Galaxies).}
Forman, W. R., 1970. Astrophys. J., 159, 719.
Isaacson, R. A., 1968. Phys. Rev., 166, 1272.
Jackson, J. C., 1970. Mon. Not. R. astr. Soc., 148, 249.
Paal, G., 1964. Acts. Phys. Hung., 17, 379.
Rood, H. J., Rothman, V. C. A. \& Turnrose, B. E., 1970. Astrophys. J., 162, 411.
Rees, M. J., 1971. Mon. Not. R. astr. Soc., 154, 187.
Sandage, A., 1961. Astrophys. J., 133, 355.
Vaucouleurs, G. de, 1971. Stars and Stellar Systems, Vol. 9, University of Chicago Press,
{\parindent=0.6cm in press.}
Weber, J., 1961. General Relativity and Gravitational Waves, Interscience, New York.
}

\end{document}